\documentclass{PoS}

\title{An application of the variational analysis to calculate the meson spectral functions}

\ShortTitle{An application of the variational analysis to calculate the meson spectral functions}

\author{\speaker{H.~Ohno}, S.~Aoki, K.~Kanaya, H.~Saito\\
        Graduate School of Pure and Applied Sciences, University of Tsukuba,
        Tsukuba, Ibaraki 305-8571, Japan \\
        E-mail: \email{ohno@het.ph.tsukuba.ac.jp}}

\author{S.~Ejiri \\
        Graduate School of Science and Technology, Niigata University,
	Niigata 950-2181, Japan \\}

\author{Y.~Maezawa \\
        Mathematical Physics Laboratory, RIKEN Nishina Center,
        Saitama 351-0198, Japan \\}

\author{T.~Umeda \\
        Graduate School of Education, Hiroshima University,
	Hiroshima 739-8524, Japan \\}

\author{(WHOT-QCD Collaboration)}

\abstract{We present a new method to calculate meson spectral functions (SPFs) on the lattice
based on a variational method. Because, on a finite volume lattice, the meson SPFs have discrete spectra only,
a suitable way to extract such discrete signals is needed.
Using a variational method, we can calculate several discrete quantities such as the 
position and the area of spectral peaks for low-lying states. Moreover data accuracy can
be improved by increasing the number of basis functions. In this report, we first confirm our method in the free quark
case and show that our method works well. Then, we apply the method to a quenched lattice QCD simulation
and calculate the charmonium SPFs for S and P-waves at zero temperature.
Our results for the ground state are well consistent with the position and the area of the lowest peaks of
charmonium SPFs calculated by the conventional maximum entropy method. For first excited states,
the signals may be reliablly extracted with our method because the charmonium mass converges to a value close to
the experimental one when the number of basis functions is increased. We also investigate the SPFs
for S-wave charmonia at below and above $T_c$. Our results suggest that $J/\psi$ and $\eta_c$ may survive
up to 1.4$T_c$.}

\FullConference{The XXVIII International Symposium on Lattice Field Theory, Lattice2010\\
		June 14-19, 2010\\
		Villasimius, Italy}

\begin{document}

\section{Introduction}
Spectral functions (SPFs) at finite temperature in quantum chromodynamics (QCD) play important
role to investigate the behavior of mesons in medium. 
Actually, charmonium SPFs are studied to understand the suppression of $J/\psi$ production
in heavy ion collision experiments such as SPS \cite{SPS} and RHIC \cite{RHIC},
which is one of the most important signals for quark-gluon-plasma formation \cite{Matsui_Satz}.
Recently, it is known that not only $J/\psi$ itself but also $\chi_c$ and $\psi'$ contribute to total yield of $J/\psi$ \cite{E705},
therefore we also need to understand the sequential $J/\psi$ suppression \cite{seq_Jpsi} due to the suppression of these charmonia.

On the lattice, the meson SPFs are extracted from Euclidean meson correlators conventionally with the maximum entropy method (MEM) \cite{MEM}
which is based on the Bayesian probability theory. According to these studies, S-wave charmonia such as $J/\psi$ and $\eta_c$ still survive up to 1.5$T_c$
in both quenched \cite{lqcd1,lqcd2,lqcd3,lqcd4} and 2-flavor QCD \cite{lqcd5}. On the other hand, P-wave charmonia such as $\chi_{c0}$ and $\chi_{c1}$
are suggested to dissolve just above $T_c$ \cite{lqcd4,lqcd5}.

For the MEM analysis, however, it is difficult to find a porper default model which shares as many properties as possible with the SPFs.
Accordingly there is ambiguity due to the choice of default model. Moreover, lattice QCD consists of only discrete spectra when spatial lattice extent is finite.
Therefore, it is important to check the conclusion drawn form MEM by other methods which can directly extract such discrete
signals instead of reproducing the continuous form of SPFs.
In this respect the variational method \cite{var} is a suitable tool to investigate a few lowest (discrete) states from Euclidean correlators.

In this study, we propose a new method to calculate the meson SPFs with the variational method.
With our method, we can determine only the value of the SPFs at discrete points corresponding to some low-lying states.
When a state dissolves, corresponding value of the SPFs should be quite modified and may become small compared with that below $T_c$.
So our goal is to investigate the temperature dependence of the values of the SPFs and find such a modification.
In the following section, we show the detail of our method and check it for free quarks.
Finally, we apply our method to the calculation of the charmonium SPFs at zero and finite temperatures in the quenched QCD.

\section{Meson SPFs with the variational method}
A meson correlator in the Euclidean space-time is defined by
\begin{equation}\label{eq:corr}
C_{\Gamma}(t) \equiv \sum_{\vec{x}}\langle\mathcal{O}_{\Gamma}(\vec{x},t)\,\mathcal{O}^{\dag}_{\Gamma}(\vec{0},0)\rangle,
\end{equation}
where $\mathcal{O}_{\Gamma}(\vec{x},t) \equiv \bar{q}(\vec{x},t)\,\Gamma \,q(\vec{x},t)$ is a meson operator and 
$\Gamma = \gamma_5,\gamma_i,\textrm{\boldmath $1$},\gamma_5\gamma_i\;(i=1,2,3)$ correspond to pseudoscalar (Ps), vector (Ve),
scalar (Sc) and axial-vector (Av) channels, respectively. For Ve and Av channels, we average the meson correlators over $i=1,2,3$.

Then meson SPFs $\tilde{\rho}_{\Gamma}(\omega)$ are given by a relation
\begin{equation}\label{eq:corr2}
C_{\Gamma}(t) = \int^{\infty}_{0} d\omega\, \tilde\rho_{\Gamma}(\omega)\,\frac{\cosh[\omega(t-N_t/2)]}{\sinh[\omega N_t/2]},
\end{equation}
where $N_t$ is the temporal lattice size.
Since we consider a finite volume system where the meson SPFs have only discrete spectra, we can write the meson SPF
as $\tilde{\rho}_{\Gamma}(\omega) = \sum_{k}\rho_{\Gamma}(m_k)\delta(\omega-m_k)$.
Therefore (\ref{eq:corr2}) is rewritten as
\begin{equation}\label{eq:corr3}
C_{\Gamma}(t) = \sum_{k} \rho_{\Gamma}(m_k)\,\frac{\cosh[m_k (t-N_t/2)]}{\sinh[m_k N_t/2]}.
\end{equation}

In order to calculate $\rho_{\Gamma}(m_k)$, we apply the variational method \cite{var} as follows.
First, we define an $n\times n$ meson correlator matrix by
\begin{equation}
{\bf C}_{\Gamma}(t) \equiv \left[C_{\Gamma}(t)_{ij}
= \sum_{\vec{x}}\langle\mathcal{O}_{\Gamma}(\vec{x},t)_i\,\mathcal{O}^{\dag}_{\Gamma}(\vec{0},0)_j\rangle \right], \quad i,j=1,\cdots,n,
\end{equation}
using smeared meson operators with the Gaussian smearing function
\begin{equation}\label{eq:sme_op}
\mathcal{O}_{\Gamma}(\vec{x},t)_i
\equiv \sum_{\vec{y},\vec{z}}\omega_i(\vec{y})\,\omega_i(\vec{z})\,\bar{q}(\vec{x}+\vec{y},t)\,\Gamma \,q(\vec{x}+\vec{z},t), \quad
\omega_i(\vec{x}) \equiv e^{-A_i |\vec{x}|^2},
\end{equation}
where $A_i$ is a smearing parameter. Here we choose $\omega_1(\vec{x}) = \delta(\vec{x})$,
namely $\mathcal{O}_{\Gamma}(\vec{x},t)_1$ is a point operator.
By solving a generalized eigenvalue problem
\begin{equation}\label{eq:gep}
{\bf C}_{\Gamma}(t)\,{\bf v}^{(k)} = \lambda_k(t;t_0)\, {\bf C}_{\Gamma}(t_0)\,{\bf v}^{(k)}, \quad k=1,2,\cdots n,
\end{equation}
effective masses $m^{\mathrm{eff}}_k(t;t_0)$ are defined by the eigenvalues as
\begin{equation}\label{eq:lambda}
\lambda_k(t;t_0) = \frac{\cosh[m^{\mathrm{eff}}_k(t;t_0)(t-N_t/2)]}{\cosh[m^{\mathrm{eff}}_k(t;t_0)(t_0-N_t/2)]}.
\end{equation}
Denoting $\Lambda = {\rm diag}\{\lambda_1,\cdots,\lambda_n\}$ and ${\bf V} = [{\bf v}^{(1)} \cdots {\bf v}^{(n)}]$,
(\ref{eq:gep}) is rewritten as ${\bf C}_{\Gamma}(t) = {\bf C}_{\Gamma}(t_0) {\bf V} \Lambda {\bf V}^{-1}$. Then we can write
the $(1,1)$ element of ${\bf C}_{\Gamma}(t)$ as
\begin{equation}\label{eq:corr4}
C_{\Gamma}(t)_{11} = \sum_{k}\left({\bf C}_{\Gamma}(t_0){\bf V}\right)_{1k}({\bf V}^{-1})_{k1}
\frac{\sinh[m^{\mathrm{eff}}_k(t;t_0)N_t/2]}{\cosh[m^{\mathrm{eff}}_k(t;t_0)(t_0-N_t/2)]}
\frac{\cosh[m^{\mathrm{eff}}_k(t;t_0)(t-N_t/2)]}{\sinh[m^{\mathrm{eff}}_k(t;t_0)N_t/2]},
\end{equation}
where $C_{\Gamma}(t)_{11}$ is point source - point sink component which equals to the meson correlator defined by (\ref{eq:corr}).
Comparing (\ref{eq:corr3}) with (\ref{eq:corr4}), we find
\begin{equation}\label{eq:spf}
\rho_{\Gamma}(m^{\mathrm{eff}}_k(t;t_0)) = \left({\bf C}_{\Gamma}(t_0){\bf V}\right)_{1k}({\bf V}^{-1})_{k1}
\frac{\sinh[m^{\mathrm{eff}}_k(t;t_0)\,N_t/2]}{\cosh[m^{\mathrm{eff}}_k(t;t_0)\,(t_0-N_t/2)]}.
\end{equation}
Here we note that we also apply the mid-point subtraction technique \cite{const_mode} in order to separate out the zero mode contribution
from the meson correlators as follows:
\begin{eqnarray*}
{\bf C}_{\Gamma}(t) \;\;\rightarrow\;\; {\bf C}_{\Gamma}(t) - {\bf C}_{\Gamma}(N_t/2).
\end{eqnarray*}
Accordingly, $\cosh$ of the kernels in (\ref{eq:corr3}), (\ref{eq:lambda}), (\ref{eq:corr4}) and (\ref{eq:spf}) is modified as
\begin{eqnarray*}
\cosh\left[m\left(t-N_t/2\right)\right] \;\;\rightarrow\;\; \cosh\left[m\left(t-N_t/2\right)\right] - 1.
\end{eqnarray*}

\section{Test in free quark case}

\begin{figure}[tbp]
 \begin{center}
  \includegraphics[width=52mm, angle=-90]{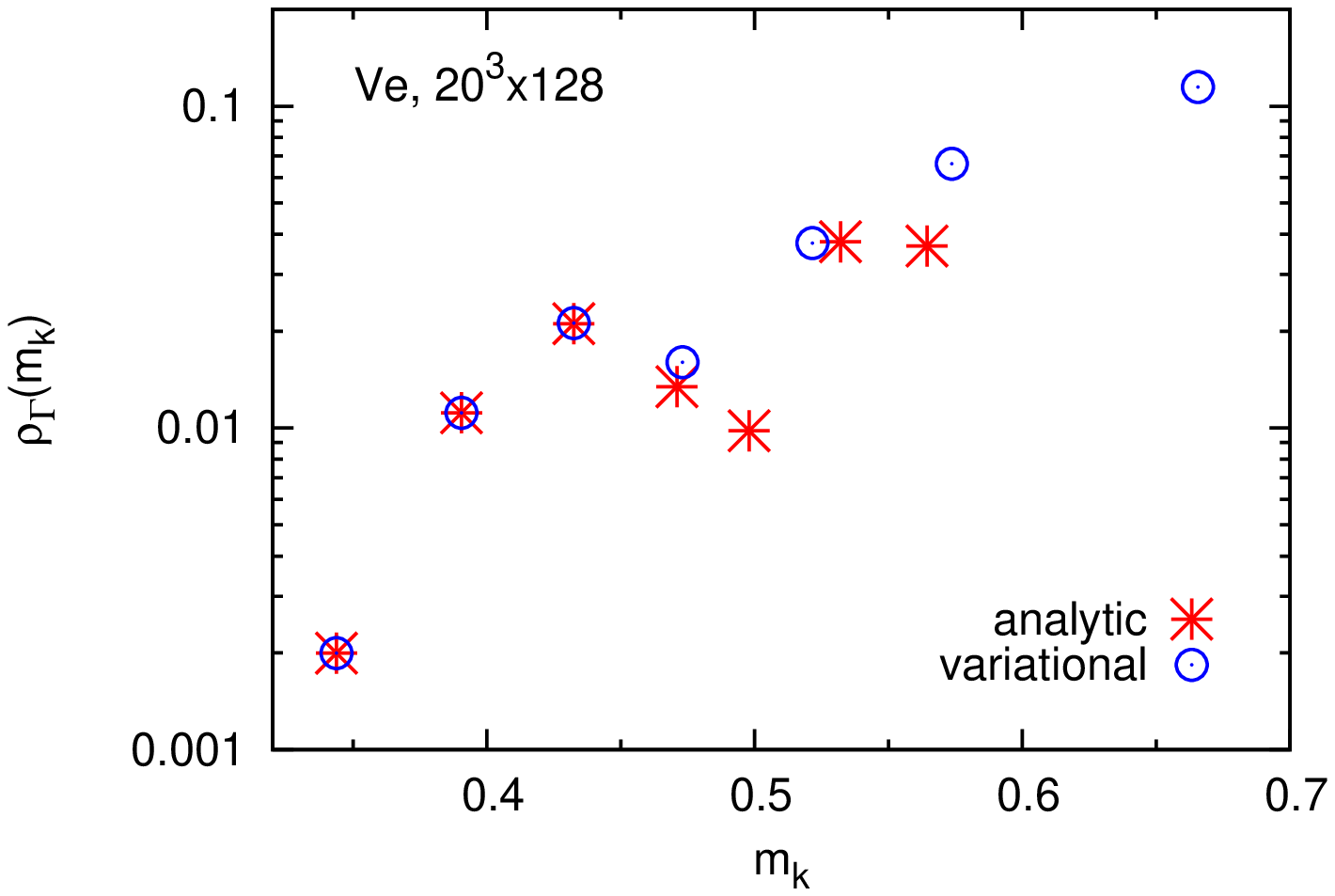}
  \includegraphics[width=52mm, angle=-90]{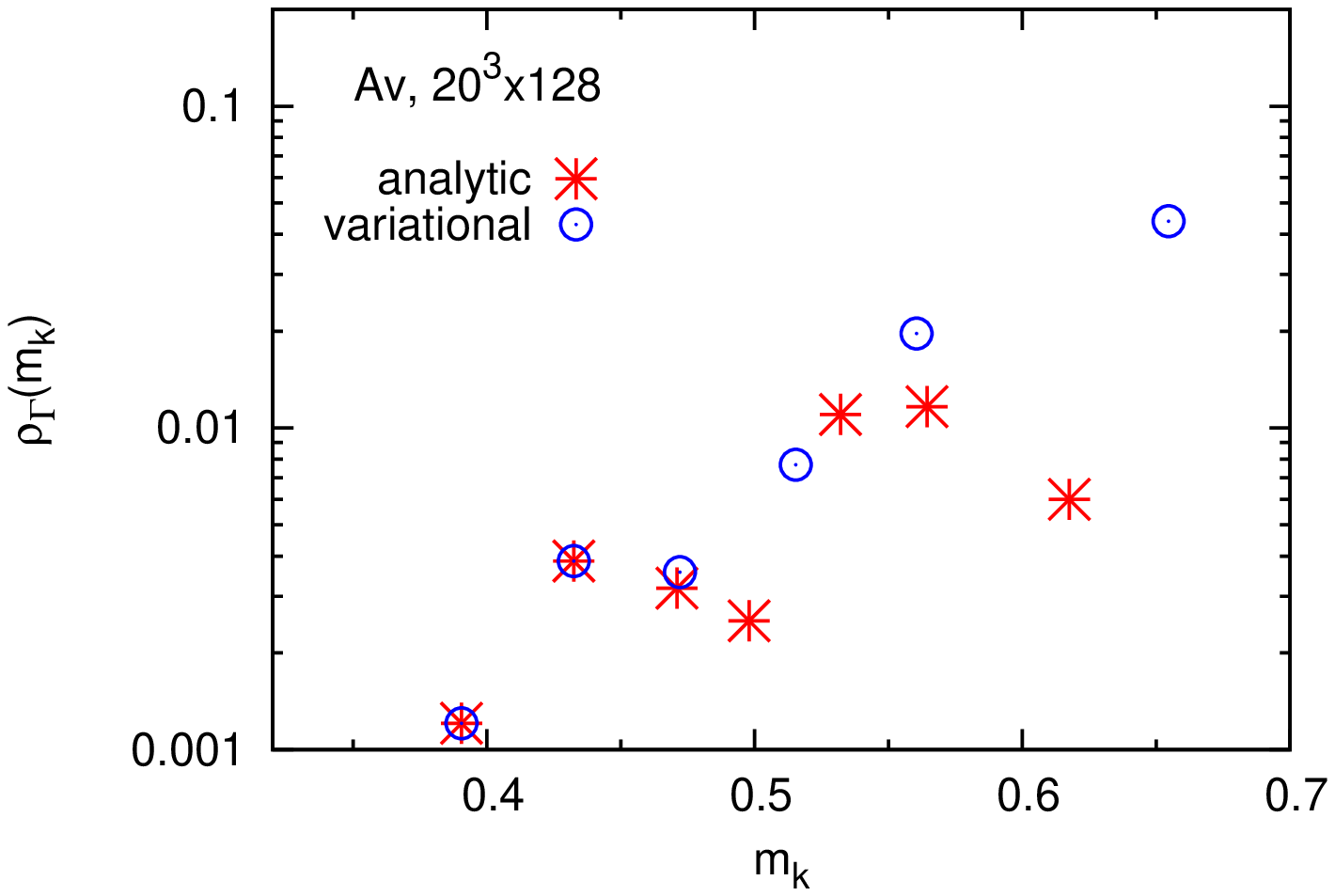}
  \caption{The meson SPF at seven lowest-lying states for Ve (left) and Av (right) channels obtained on a $20^3\times 128$ lattice in the free Wilson quark case. 
  Circle symbols are the results of the variational method for $n=7$ and the analytic solutions are also shown by the asterisks.}
  \label{free_SPF}
 \end{center}
\end{figure} 

First, in order to confirm our method, we consider the free Wilson quark case.

We study on an anisotropic lattice with the anisotropy $\xi \equiv a_s/a_t=4$
where $a_s$ and $a_t$ are the spacial and temporal lattice spacings, respectively. We choose $m_{0}a_s \approx 0.7501$ as the quark mass so that
the ground state meson masses studied in the next section are approximately reproduced. We also choose $r=1$ as the Wilson parameter.

For the smeared meson operators defined in (\ref{eq:sme_op}), $n=7$ smearing parameters are chosen as $A_1=\infty$, $A_2=0.25$, $A_3=0.20$, $A_4=0.15$,
$A_5=0.10$, $A_6=0.05$ and $A_7=0.02$, where $A_1=\infty$ corresponds to the point operator.
Then we construct $n \times n$ meson correlator matrices on $20^3\times 128$ lattice and calculate $m_k$ and $\rho_{\Gamma}(m_k)$ with the variational method
for Ps, Ve, Sc and Av channels. Since we do not have so much space, we show the results only for Ve and Av channels in Figure \ref{free_SPF}.
We choose $t_0=51$ and $t=63$ where $t_0$ and $t$ are as large as possible keeping the signals stable.
The analytic solutions are also shown.

This figure shows that the results are almost consistent with the analytic solutions up to the second excited state for each channel.
Here it is noted that it looks more difficult to extract higher states' signals for P-wave (Av) than S-wave (Ve).
The results for Ps and Sc channels are almost the same as those for Ve and Av channels, respectively.
Therefore it is shown that we can calculate the meson SPFs at some low-lying states with the variational method using several basis operators.

\section{Charmonium SPFs}
Next, we apply our method to the quenched QCD and calculate the charmonium SPFs.

Our simulations are preformed on $20^3\times N_t$ anisotropic lattices with the renormalized anisotropy $\xi=4$. Adopting the standard plaquette gauge
action, the gauge coupling and the bare gauge anisotropy parameter are chosen $\beta=6.10$ and $\gamma_G=3.2108$.
Then the lattice spacing $a_s=0.0970(5)$ fm ($a^{-1}_s=2.030(13)$ GeV) is determined by the Sommer scale $r_0=0.5$ fm \cite{Sommer}, which means that
our spacial volume is about (2 fm)$^3$. For the temporal lattice size, we adopt $N_t=160$ at zero temperature and $N_t=32,\,26,\,20$ at about $0.88T_c$, $1.1T_c$,
$1.4T_c$, respectively, where the critical temperature $T_c$ determined by peak position of the Polyakov loop susceptibility corresponds to $N_t \approx 28$.
After 20,000 pseudo-heat-bath sweeps for thermalization, separated by 500 sweeps, 299 and 800 configurations are generated at zero and finite temperature, respectively.
For valence quarks, we adopt $O(a)$-improved Wilson quark action with the bare quark anisotropy parameter $\gamma_F=4.94$ and tree-level tadpole improved clover
coefficients $c_E=3.164$ and $c_B=1.911$. The definition of the actions and their parameters are almost the same as those of Ref. \cite{action}, however we choose different
Wilson parameter $r=1$ to suppress lattice artifacts in higher excited charmonia \cite{Karsch}. 
In order to roughly reproduce the experimental value of $J/\psi$ mass, $\kappa=0.10109$ is chosen
where this value of $\kappa$ corresponds to the charm quark mass on an isotropic lattice with a similar spatial lattice spacing.

\begin{figure}[tbp]
 \begin{center}
  \includegraphics[width=52mm, angle=-90]{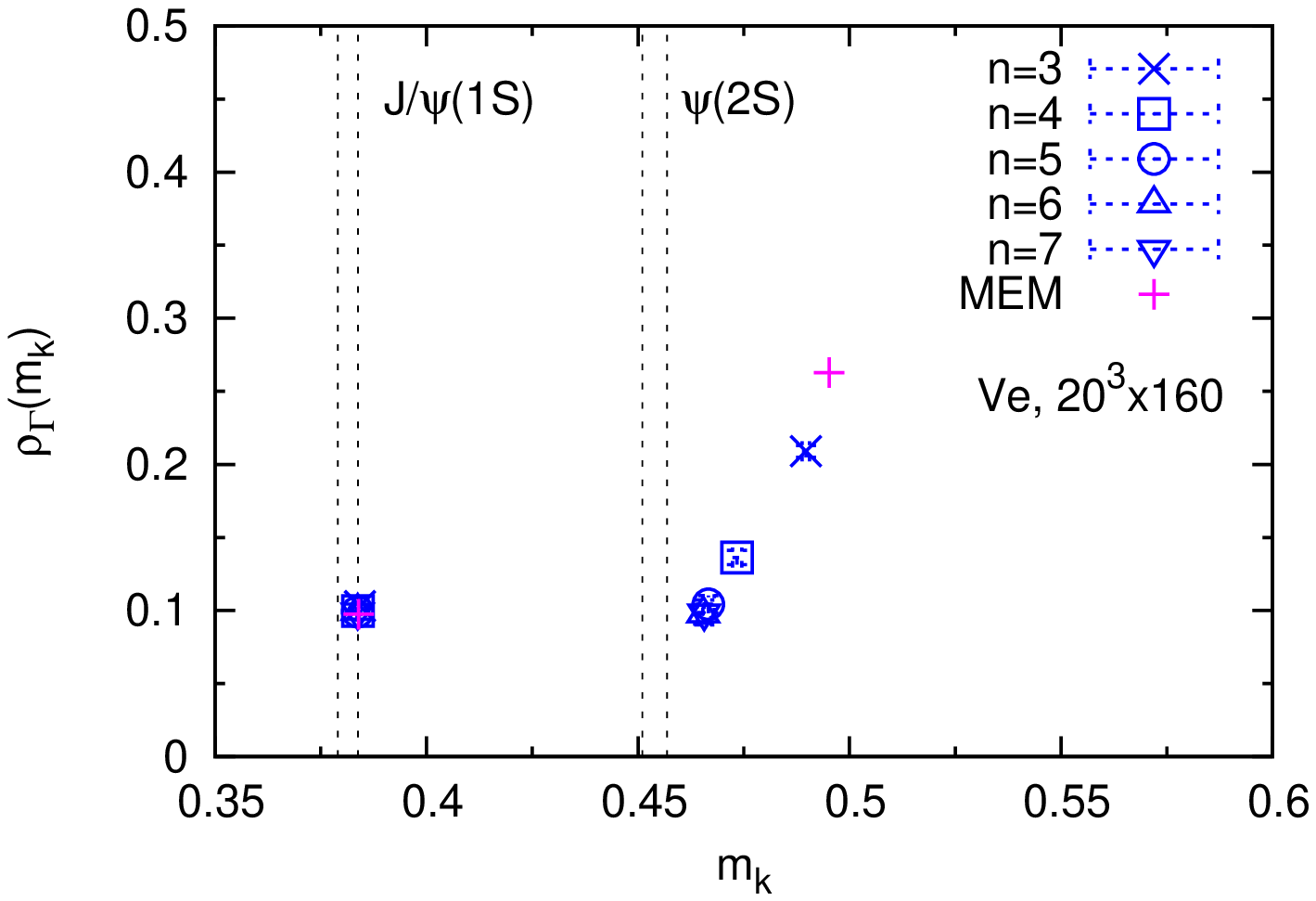}
  \includegraphics[width=52mm, angle=-90]{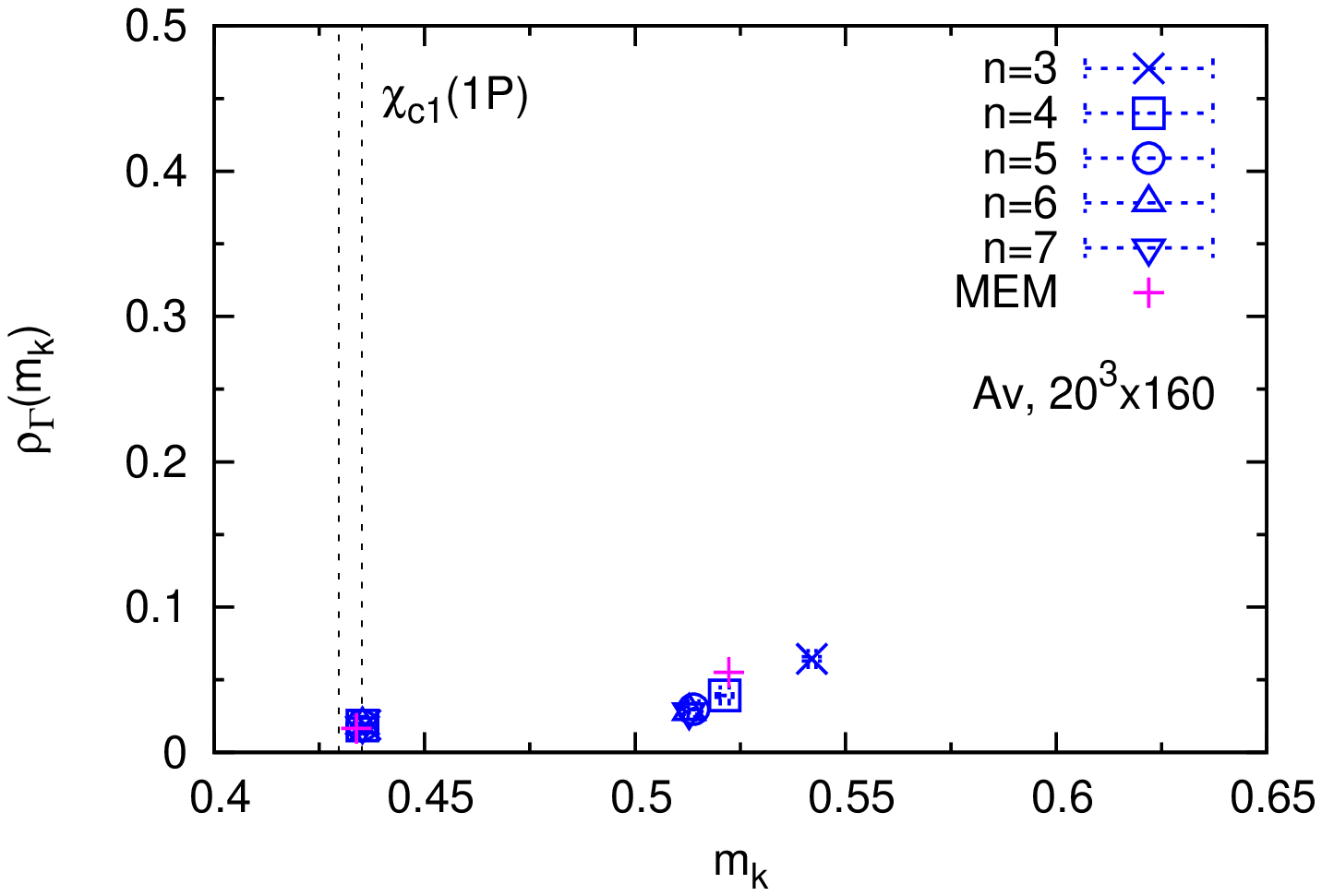}
  \caption{Charmonium SPF at the ground and the first excited states for Ve (left) and Av (right) channels at zero temperature obtained on the $20^3\times 160$ lattice.
  Cross, square, circle, triangle and reverse triangle symbols indicate the data by the variational method with $n=3,\,4,\,5,\,6,\,7$, respectively,
  and plus symbols indicate the MEM data. The vertical dashed lines indicate the range of experimental masses for $J/\psi({\rm 1S})$ and $\psi({\rm 2S})$ mesons
  for Ve channel and $\chi_{c1}(1P)$ meson for Av channel \cite{PDG2010}.}
  \label{var_mem}
 \end{center}
\end{figure} 

First, we calculate $m_k$ and $\rho_{\Gamma}(m_k)$ for the ground and the first excited states for Ps, Ve, Sc and Av channels
at zero temperature with the variational method.
We use the same smearing parameters $A_1, A_2, \cdots, A_n$ introduced in previous section and vary the number of basis operators as $n=3,\,4,\,5,\,6,\,7$.
$t_0=5$ is chosen as the reference point and the plateau is fitted with the range of $t=73 - 77$ and $t=35 - 39$ for the ground state and the first excited state
of S-wave, respectively, and $t=60 - 64$ and $t=25 - 29$ for the ground state and the first excited state of P-wave, respectively.
We have reasonable $\chi^2/$d.o.f which is from 0.1 to 5.0 in this fit analysis.

To compare our results with those given by the conventional method, we also calculate the charmonium SPFs for the point operators with MEM
and regard position and area of the each peak as $m_k$ and $\rho_{\Gamma}(m_k)$, respectively. Our MEM analysis basically follows Ref. \cite{MEM}.
As the default model function $m(\omega)$, we use a form $m(\omega)=m_{DM}\omega^2$ where $m_{MD}=4.2$ for Ps and Sc channels and $m_{MD}=2.4$ for
Ve and Av channels, which are determined according to the asymptotic behavior of the meson correlators in perturbation theory \cite{MEM, lqcd1}.
The charmonium correlators at $t=1 - 60$ and $t=3 - 60$ are used for S-wave and P-wave, respectively, since those near $t=0$ are suffered from lattice artifacts.
Furthermore we check other basic parameters of MEM and confirm the results are stable around our parameters.
The peak positions are defined at the maxima of the SPF. To calculate the area of peaks, we divide the SPFs into each peak at minima of the SPF
when each peak of the SPFs are not isolated. 

Figure \ref{var_mem} shows the results for Ve and Av channels. The range for the experimental values of charmonium masses \cite{PDG2010}
are also shown by the vertical dashed lines. 

For the ground state of these channels, the data for the variational method for each $n$ and that for MEM are well consistent with each other and close to
the experimental value. On the other hand, the first excited state's data are quite different from each other. However, for S-wave,
the data for the variational method are converged on a certain point close to the experimental value as $n$ increases. This means that,
in our method, the quality of the signals for higher states can be improved by increasing the number of basis operators.
The results for Ps and Sc channels are almost the same as those for Ve and Av channels, respectively.

\begin{figure}[htbp]
 \begin{center}
  \includegraphics[width=52mm, angle=-90]{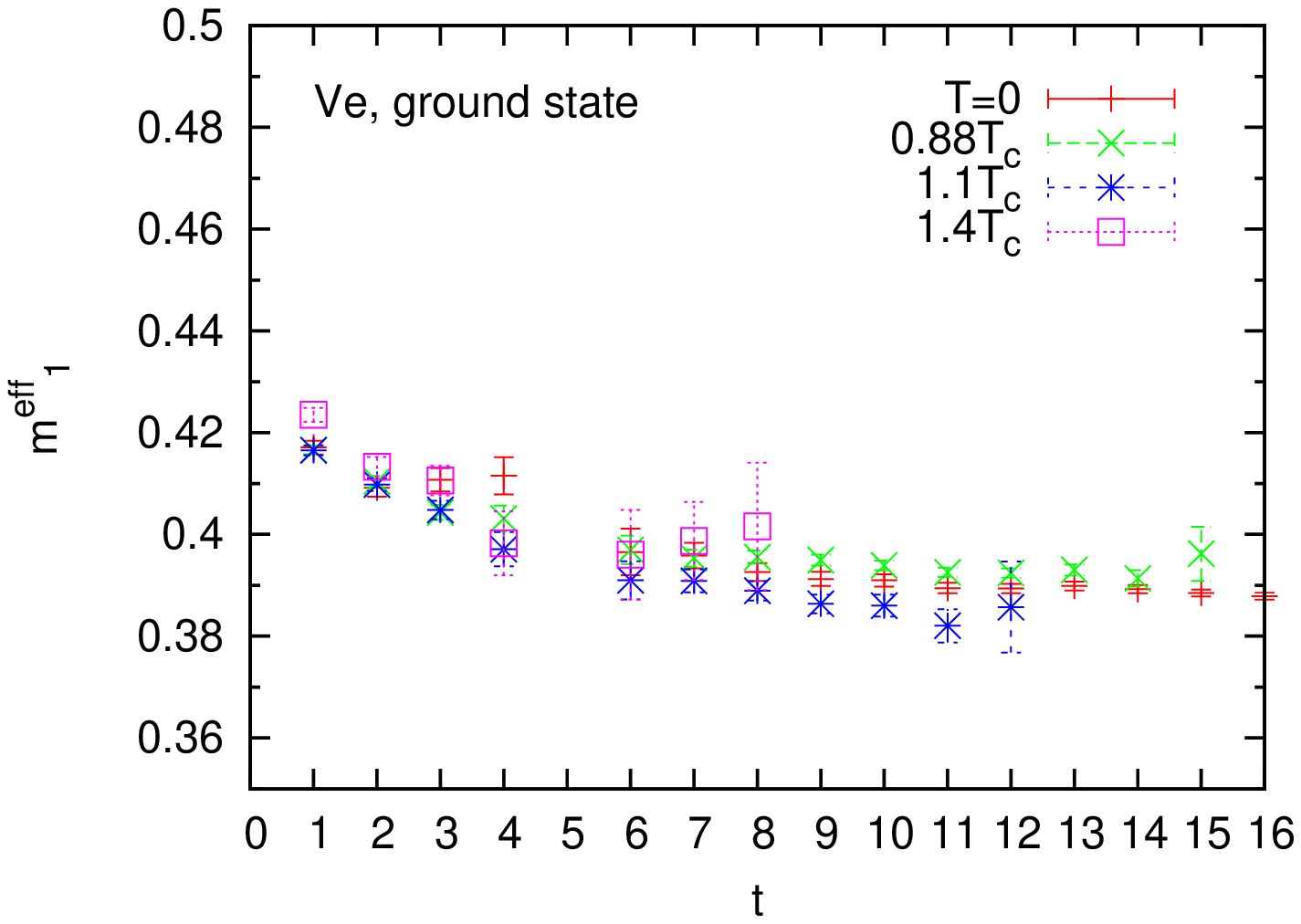}
  \includegraphics[width=52mm, angle=-90]{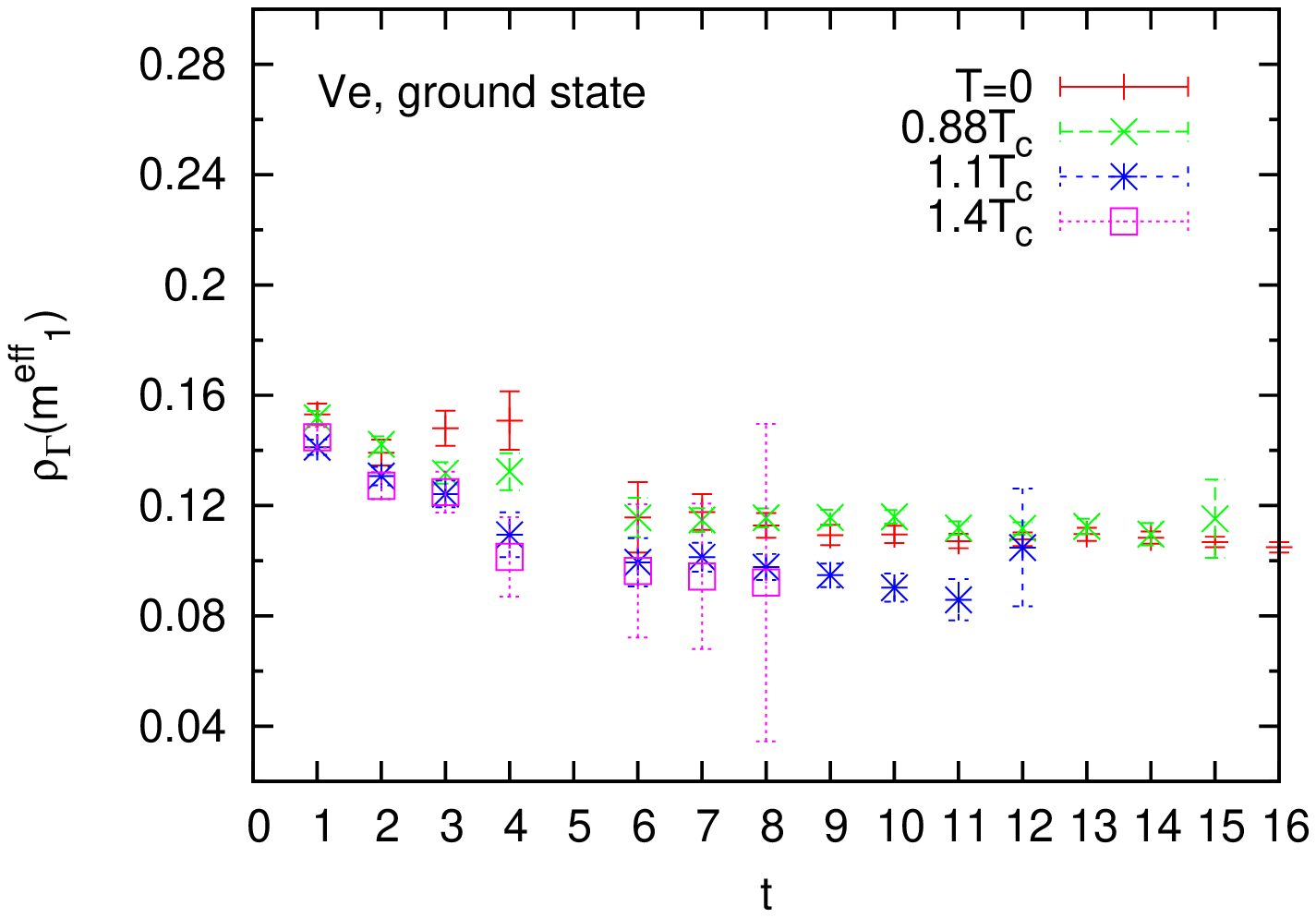}
  \caption{The temperature dependence of the effective mass (left) and corresponding value of the SPF (right)
  for the ground state of Ve channel. The plus, cross, asterisk, square symbols indicate the data at $T=0,0.88T_c,\,1.1T_c,\,1.4T_c$, respectively.}
  \label{temp_dep_Ve}
 \end{center}
\end{figure} 

Finally, we investigate the temperature dependence of the effective masses and corresponding values of the SPFs
for the ground state of S-wave charmonia with the variational method using the same smearing parameters as those at zero temperature
and choosing $n=7$.
Figure \ref{temp_dep_Ve} shows the results for Ve channel at $t_0=5$ as a function of $t$ up to 1.4$T_c$. The mass of the ground state and corresponding
value of the SPF can be approximately obtained by a constant fit of the plateau region.

It seems that the effective mass has no clear temperature dependence. On the other hands,
there may be some difference between the value of the SPF below $T_c$ and that above $T_c$, although the difference is quite small.
The result for Ps channel is almost the same as that for Ve channel. 
We find no clear evidence for dissociation of $J/\psi$
and $\eta_c$ up to 1.4 $T_c$.

\section{Conclusions}
We proposed a new approach to calculate the meson SPFs using the variational method.
To test our method, we calculated the meson SPFs for Ps, Ve, Sc, and Av channels in the free quark case. 
Using 7 basis operators  defined by a Gaussian smearing function, we find that 
the results are well consistent with the analytic solutions up to the second excited state.

At zero temperature, we calculated the charmonium SPFs for S and P-wave up to the first excited states varying the number of the basis operators.
Comparing our results with those given by MEM, we find that the data for the ground state is consistent with each other.
On the other hand, the data for the first excited state are discrepant with each other. However, 
the signals in our method can be improved by increasing the number of basis operators.

At finite temperature, we investigated the temperature dependence of the effective masses and corresponding values of the SPFs
for the ground state of S-wave charmonia.
We find no clear evidences of dissociation for $J/\psi$ and $\eta_c$ mesons up to 1.4$T_c$.

\vspace{5mm}
This work is in part supported by Grants-in-Aid of the Japanese Ministry of Education, Culture, Sports, Science and Technology, 
(Nos.~20340047, 21340049, 22740168, 22840020) and by the Grant-in-Aid for Scientific Research on Innovative Areas (No. 2004: 20105001, 20105003).
The simulations have been performed on a supercomputer NEC SX-8 at the Research Center for Nuclear Physics (RCNP) at Osaka University.

\end{document}